# A COMPUTATIONAL ANALYSIS OF ORAL ARGUMENT IN THE SUPREME COURT

*Gregory M. Dickinson\**

## INTRODUCTION

Term after term the scene is the same. Before the Supreme Court announces its opinions, we carefully study the oral argument transcripts, pore over the questions asked by each Justice, and tally the number of questions asked to each side, all hoping to find some clue about how the Justices might vote. Within days or even hours of oral argument, court watchers have carefully deconstructed the arguments and highlighted the question-answer exchanges they deem most telling of the Justices' views.

---

\* Senior Associate, Harter Secrest & Emery LLP; Fellow, Northwestern Pritzker School of Law (2016–2018); J.D., Harvard Law School; B.S., Computer Science, Houghton College. Thanks to Michael Heise, Jason Iuliano, Ben Johnson, Jim Lindgren, John McGinnis, and participants in the 2017 Junior Faculty Law & Technology Workshop for their helpful comments on early drafts. A special thank you to my brother, Brian Dickinson, who shares my peculiar passion for computational study of the law, and who has been a ready and willing sounding board far beyond any fraternal obligation.

Immediate media coverage followed even a case like *Yates v. United States*, which posed the question whether tossing undersized red grouper overboard to impede a fish and wildlife investigation constituted the destruction of "records, documents, or tangible objects" within the meaning of the Sarbanes-Oxley Act. Fascinating as the case might be to legal academics, one might have been forgiven for thinking it would fail to capture the attention of the broader public. Yet, within a day of argument, the *New York Times* published a detailed article reporting on the case and zeroing in on the following delightful exchange to highlight Chief Justice Roberts's skepticism of the government's case.[1]

> CHIEF JUSTICE ROBERTS: Well, what if you stopped [a person] on the street and said is a fish a record, document or tangible object?
> MR. MARTINEZ: I think if you—if you asked them that question and you—you pointed them to the fact that—
> JUSTICE SCALIA: I don't think you would get a polite answer.[2]

The depth and rapidity of post argument analysis is astounding. Oral argument has long since been displaced by the written brief as the primary means by which advocates present their arguments to the Court, but the institution remains the most public, most accessible, and perhaps most scrutinized component of judicial decision making.

Yet the basic function and operation of oral argument as an institution is not well understood. Political scientists and legal scholars continue to dispute fundamental questions of how legal doctrine, political ideology, judicial psychology, and other factors converge to produce judicial decisions.[3] The role of oral arguments in the process is even less clear. Some emphasize oral argument's usefulness as an information-gathering tool; others discuss its role as a forum for judges to form and test their views and build voting coalitions; and still others attribute little significance to the institution at all.[4]

---

[1] Adam Liptak, *Justices Consider Whether Tossing Out Fish Destroyed Records*, N.Y. TIMES, Nov. 6, 2014, at A20.

[2] Transcript of Oral Argument at 47, Yates v. United States, 135 S. Ct. 1074 (2015) (No. 13-7451).

[3] *See, e.g.*, JEFFEREY A. SEGAL & HAROLD J. SPAETH, THE SUPREME COURT AND THE ATTITUDINAL MODEL REVISITED 86 (2002) (presenting the attitudinal model of judicial decision making, which explains judicial decisions as products of judges' preferred social policies); LEE EPSTEIN, WILLIAM M. LANDES & RICHARD A. POSNER, THE BEHAVIOR OF FEDERAL JUDGES: A THEORETICAL & EMPIRICAL STUDY OF RATIONAL CHOICE 124–37 (2013) (observing judicial ideological to be an important and growing factor in judicial decision making, but noting that legal commitments such as stare decisis also play an important role, particularly in less salient cases); MICHAEL A. BAILEY & FORREST MALTZMAN, THE CONSTRAINED COURT: LAW, POLITICS, AND THE DECISIONS JUSTICES MAKE 2–9 (2011) (explaining judicial decision making as the product of both individual judicial preferences and various "constraints" including legal doctrine and political pressures such as potential rebuke by the executive and legislative branches).

Past study of oral argument has tended to proceed in one of two directions: first, focused review by human readers of arguments in a small number of cases; and second, empirical study of quantifiable attributes across all cases. Tea-leaf-reading post argument analysis of Justices' likely votes falls into the first category, as do studies that rely on personal observation or review of a limited sample of cases to identify the various functions of oral argument[5] or assess the Justices' individual communication styles.[6] Falling in the second group, for example, are efforts to correlate ultimate votes and case outcomes with heavy or light questioning,[7] especially wordy questioning,[8] inclusion in questions of an unusual number of negative or positive words,[9] audio and facial attributes of Supreme Court advocates,[10] and the frequency with which the Justices interrupt one another as they compete to question the advocates.[11] Also falling into this category are those studies that have examined long-term trends in inter-Justice and Justice-advocate communication dynamics, such as the Justices' increasing use of non-questions to rebut colleagues or advocate for favored parties,[12] and discrepancies in question interruptions at oral argument that are attributable to the gender, ideology, and seniority of the advocates and Justices.[13]

---

[4] *See generally* TIMOTHY R. JOHNSON, ORAL ARGUMENTS AND DECISION MAKING ON THE UNITED STATES SUPREME COURT 23–24 (2004) (positing that judges use oral argument time to gather the information necessary to decide cases in accord with their policy preferences); RYAN A. MALPHURS, RHETORIC AND DISCOURSE IN SUPREME COURT ORAL ARGUMENTS: SENSEMAKING IN JUDICIAL DECISIONS 38–46, 86–92 (2013) (discussing oral argument as part of a judicial cognitive "sensemaking" process in which judges attempt to situate the case among their existing intellectual commitments, advance and test the strength of their views, influence colleagues, and form voting coalitions); SEGAL & SPAETH, *supra* note 3, at 280 (suggesting that oral argument has little impact on case outcomes).

[5] *See, e.g.*, MALPHURS, *supra* note 4, at 60–101 (analyzing oral arguments from three leading cases and employing psychology and communications theory methodologies to identify the functions of oral argument and highlight the Justices' individual approaches).

[6] *See, e.g.*, LAWRENCE S. WRIGHTSMAN, ORAL ARGUMENTS BEFORE THE SUPREME COURT: AN EMPIRICAL APPROACH 85–104 (2008) (classifying each Justice's questioning style based on five traditional personality factors: extraversion, agreeableness, conscientiousness, openness to experience, and negative affectivity).

[7] *See* Timothy R. Johnson et al., *Inquiring Minds Want to Know: Do Justices Tip Their Hands with Questions at Oral Argument in the U.S. Supreme Court?*, 29 WASH U. J. L. & POL'Y 242 (2009).

[8] EPSTEIN ET AL., *supra* note 3, at 317–24 (finding the average number of words per question directed to a party to be negatively correlated with that party's likelihood of success).

[9] *See* Ryan C. Black et al., *Emotions, Oral Arguments, and Supreme Court Decision Making*, 73 J. POL. 572 (2011) (use of pleasant or unpleasant language toward an advocate correlated with likelihood of success).

[10] Daniel L. Chen et al., *Is Justice Really Blind? And is it Also Deaf?* (July 31, 2016), *available at* https://ssrn.com/abstract=2816567.

[11] *See* Tonja Jacobi & Kyle Rozema, *Judicial Conflicts and Voting Agreement: Evidence from Interruptions at Oral Argument*, 59 B.C. L. REV. 2259 (2018).

Each approach has its strengths. Empirical studies focusing on a small number of discrete, quantifiable attributes such as the number of questions asked to each advocate or the appointing president of each Justice allow broad generalizations about oral argument and judicial decision making: Justices tend to vote in accordance with their ideological preferences, and they tend to ask more questions when they are skeptical of a party's position.[14] Studies based on expert human observation of oral argument, in turn, add anecdotal evidence regarding the Justices' questioning behaviors, but lack the rigor and generalizability of a large-scale empirical study.[15]

Despite their important contributions, these approaches leave another important facet of oral argument unexplored: the dialogue between Justice and advocate that is at the heart of the institution. This study relies on machine learning techniques to, for the first time, construct predictive models of judicial decision making based not on oral argument's superficial features or factors external to oral argument, such as where the case falls on a liberal-conservative spectrum, but on the actual content of the oral argument itself—the Justices' questions to each side. The resultant models offer an important new window into aspects of oral argument that have long resisted empirical study, including the Justices' individual questioning styles, how each expresses skepticism, and which of the Justices' questions are most central to oral argument dialogue.

## I. The Data

Until 2004, Supreme Court oral argument transcripts did not identify individual Justices by name. Instead, transcripts introduced all statements from the bench simply as "Question." The Supreme Court abandoned this long-standing practice in 2004 and began, for the first time, to associate particular Justices with particular questions. Official transcripts of Supreme Court arguments are available on the Court's website in PDF format. Fourteen terms of Justice-identified transcripts are now available. In addition, the Oyez Project at Chicago-Kent College of Law has analyzed audio recordings of oral arguments and employed voice-recognition software to match questions with particular Justices in oral argument transcripts for an additional six years, dating back to 1998.

---

[12] *See* Tonja Jacobi & Matthew Sag, *The New Oral Argument: Justices as Advocates*, 94 Notre Dame L. Rev. (forthcoming 2019), *available at* https://papers.ssrn.com/sol3/papers.cfm?abstract_id=3125357.

[13] *See* Tonja Jacobi & Dylan Schweers, *Justice, Interrupted: The Effect of Gender, Ideology, and Seniority at Supreme Court Oral Arguments*, 103 Va. L. Rev. 1379 (2017).

[14] *See* Johnson, *supra* note 4, at 13–15; Segal & Spaeth, *supra* note 3, at 312.

[15] *See* Johnson, *supra* note 4, at 14.

This project incorporates transcript data from both sources. The dataset is derived from oral argument transcripts of the nearly 1,400 cases argued and decided from 1998 to 2015, a period spanning the tenure of two Chief Justices and eleven Associate Justices.[16] Transcript data were paired with case outcome and Court composition data from the Supreme Court Database (SCDB), a widely used source of expertly coded data on the Supreme Court.

To prepare the transcripts for processing, PDF and HTML transcript files were converted to a JSON file format that includes basic case data and pairs each utterance with a particular speaker. The files record for each argument the docket number and case name, the content of each statement, the speaker who uttered it, to whom, whether the speaker was a Justice or an advocate, and, if an advocate, for which side.

## II. Feature Extraction

Oral argument holds a treasure trove of information regarding the Justices' attitudes toward a case. Human observers can take cues from the tone of a Justice's voice and infer likely leanings from the phrasing of a tricky question or an imbalance in questions directed to one side or the other. A knowledgeable observer's subject-matter expertise also allows him or her to follow along as the Justices employ questions to situate the present case within a sizeable body of previous case law that, even if never discussed, lurks beneath the surface of each argument.

Computer models, of course, cannot intuitively detect a tricky or skeptical question phrasing and lack the subject-matter expertise to understand a question's relationship to prior case law. To build a predictive voting model from oral argument transcripts, the information contained in those transcripts must be extracted into data points, "features," that serve to represent the content of the oral argument in numerical terms susceptible to computational analysis.

This project relies on four categories of such features: Question Count Features, Question Chronology Features, Question Sentiment Features, and N-Gram Features. These features numerically represent the volume, timing, friendliness or unfriendliness, and content of the Justice's questions to each side during oral argument.

---

[16] This eighteen-year period begins during the Second Rehnquist Court, a long period of stability in the Court's composition, and continues through the final full term of cases heard and decided by Justice Scalia. It does not include cases argued before the Court's most recent appointees, Justices Gorsuch and Kavanaugh, for whom little case data is available.

A. *Question Count Features*

Question Count Features describe the frequency and length of the Justices' questions to each side.

| Question Count Features |
| --- |
| NumQuestionsToPetitioner |
| NumQuestionsToRespondent |
| AveWordsPerQuestionToPetitioner |
| AveWordsPerQuestionToRespondent |
| PercentQuestionsToPetitioner |
| PercentQuestionsToRespondent |

Work by Timothy Johnson, Lee Epstein, and others has shown that the more questions a judge asks of a party at oral argument, the more likely she is to vote against that party and the more likely that party is to lose.[17] This is especially true when the number of questions a party receives outnumber those directed to his opponent.[18]

Judicial behaviorists posit two possible theories for this phenomenon.[19] The first, the legalistic view, is that judges direct more questions to the side they disfavor as a natural part of their decision-making processes. Having read the parties' briefs and perceived one side's position as weak, or having had an initial question unsatisfactorily answered, a judge may test that side's attorney with probing questions to confirm or disconfirm her initial evaluation. By contrast, under the second view, the realist view, the question imbalance is primarily attributable to the dynamics of judicial deliberation in multijudge panels. Judges may direct more questions to the side they disfavor to expose the weakness in that side's case. They may also use questioning as an indirect mechanism for communicating their own views to their colleagues—perhaps in less delicate terms than would be appropriate directly.

However one explains the phenomenon, the number of questions the Justices ask at oral argument is an important measure of their attitudes toward a case. The Question Count Features capture the frequency, length, and percentage of questioning directed to each party. Because the correlation between question counts and case outcomes is well established,[20] the Question Count Features also provide an important baseline

---

[17] *See* Johnson et al., *supra* note 7, at 156–59; EPSTEIN ET AL., *supra* note 3, at 317–24; *see also* John G. Roberts, Jr., *Oral Advocacy and the Re-emergence of a Supreme Court Bar*, 30 J. SUP. CT. HIST. 68, 75 (2005).

[18] *See* Johnson et al., *supra* note 7, at 249–50; EPSTEIN ET AL., *supra* note 3, at 316–17; *see also* Roberts, Jr., *supra* note 17.

[19] For an in-depth discussion and application of the legalistic and realist views of judicial behavior, see EPSTEIN ET AL., *supra* note 3, at 305–11.

[20] *See* Johnson et al., *supra* note 7, at 156–59; EPSTEIN ET AL., *supra* note 3, at 317–24; *see also* Roberts, Jr., *supra* note 17, at 75.

from which to assess the marginal predictive value of other features discussed below.

## B. Question Chronology Features

The Question Chronology Features measure how quickly a particular Justice began questioning each side and whether the Justice's engagement with each side was extended over multiple consecutive questions.

| *Question Chronology Features* |
| --- |
| FirstQuestionToPetitioner |
| FirstQuestionToRespondent |
| AveConsecutiveQuestionsToPetitioner |
| AveConsecutiveQuestionsToRespondent |

Advocates before the Supreme Court prepare their arguments knowing that they are unlikely to get in more than a few minutes of remarks before one of the Justices interrupts with a question.[21] This trend has only accelerated over the last few decades: The amount of oral argument time consumed by the Justices' questions has nearly doubled since 1979, leaving the advocates less and less time to make their own arguments as the Justices jockey for position, interrupt one another, and play to the crowd.[22]

Given the scarcity of oral argument time and the Court's reputation for vigorous questioning, a Justice's unusually sustained or especially early questioning may signal that the Justice has serious doubts about a party's case. The Question Chronology Features attempt to capture that information in two sets of features. The *FirstQuestion* features indicate how many questions into the argument a particular Justice asked his first question. The *AveConsecutiveQuestions* features indicate the average number of consecutive questions a Justice asked a party without another Justice interrupting to ask a question.

## C. Question Sentiment Features

Question Sentiment Features indicate the sentiment of each Justice's questions to each side.

| *Question Sentiment Features* |
| --- |
| AveSentimentToPetitioner |
| AveSentimentToRespondent |

---

[21] *See* EPSTEIN ET AL., *supra* note 3, at 313; *see also* Adam Liptak, *Nice Argument, Counselor, But Let's Hear Mine*, N.Y. TIMES, Apr. 5, 2011, at A12.

[22] *See* EPSTEIN ET AL., *supra* note 3, at 311–12; *see also* Liptak, *supra* note 21.

A growing body of literature documents various ways that the Justices' conscious and subconscious emotional responses can telegraph their views of a case.[23] Ryan Black and others have shown, for example, that the number of "pleasant" versus "unpleasant" words in questioning at oral argument correlates with case outcomes, with a party's probability of winning jumping by about 20% when the Court directs more unpleasant language to his or her opponent.[24] Similarly, recent work analyzing audio recordings of oral argument has shown that further information can be gleaned by considering the Justices' vocal pitch, which tends to rise in moments of high emotional arousal.[25]

For this project, the sentiment of each Justice's questions to each side was measured using the Stanford CoreNLP sentiment annotator, an open-source sentiment analysis tool that evaluates the sentiment of a given text on a scale ranging from very negative to very positive (1–5). The CoreNLP sentiment annotator breaks sentences into their parts of speech and interprets the function of each word in the sentence before evaluating the sentiment of a sentence as a whole.

The CoreNLP annotator's authors offer the following movie review as an example of the benefits of this approach: "This movie was actually neither that funny, nor super witty."[26] To a human reader, the review is plainly critical; the implication is that the movie attempted but failed to be witty and funny. But raw counts of positive versus negative words give the impression that the review is positive because it includes the strongly positive words, "funny," "super," and "witty."

The review's true sentiment can only be understood by analyzing the structure of the sentence and taking into account two key negations that affect the meaning of the positive words that follow. The figure below shows how the CoreNLP annotator divides the sentence into its component parts and weights the entire sentence as negative because the positive predicate adjectives modifying the subject are preceded by negations.

---

[23] *See* Chen et al., *supra* note 10, at 1 (commenting on existing body of literature correlating emotional arousal with case outcomes and building on it to construct a predictive model based on facial attributes).

[24] Black et al., *supra* note 9, at 577 (finding correlation between question pleasantness and case outcomes over 1979 to 2008 study period).

[25] *See* Bryce J. Dietrich et al., *Emotional Arousal Predicts Voting on the U.S. Supreme Court* POL. ANALYSIS 1-7 (Nov. 8, 2018), https://scholar.harvard.edu/files/msen/files/scotus-audio.pdf.

[26] *See* Deeply Moving: Deep Learning for Sentiment Analysis, http://nlp.stanford.edu/sentiment/. An online demo of the tool is available on the Stanford NLP website at http://nlp.stanford.edu:8080/sentiment/rntnDemo.html. For a complete description of the algorithm, see Richard Socher et al., *Recursive Deep Models for Semantic Compositionality Over a Sentiment Treebank*, 1631 CONF. ON EMPIRICAL METHODS IN NAT. LANGUAGE PROCESSING (2013).

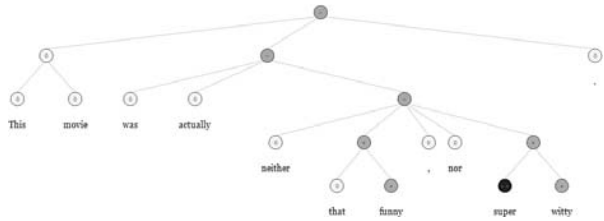

This approach permits more nuanced sentiment analysis than the pioneering work of Black et al., which relied on a word-sentiment dictionary to tally the number and intensity of positive versus negative words in a sentence, but did not take into account each word's role in the sentence.[27]

### D. N-Gram Features

N-Gram Features quantify the content of the Justices' questions to each side. A human listener will recognize important subject matter and vocabulary patterns in the Justices' questions at an intuitive, even subconscious level. A question beginning "but, counsel, you can't tell me . . ." tips off the listener that the Justice is skeptical before the Justice even finishes his question. N-Gram Features capture these sorts of question characteristics by recording counts of the number of times a Justice uses a particular word or phrase in his or her questions to a party. The counts can later be compared with Justice votes for or against that party to identify whether the n-gram is positive, negative, or neutral.

To take an example, the phrase above, "but, counsel, you can't tell me," contains fifteen distinct n-grams between one and three words in length.

| | |
|---|---|
| Full phrase | but counsel you can't tell me |
| Unigrams ($n=1$) | "but"; "counsel"; "you"; "can't"; "tell"; "me" |
| Bigrams ($n=2$) | "but counsel"; "counsel you"; "you can't"; "can't tell"; "tell me" |
| Trigrams ($n=3$) | "but counsel you"; "counsel you can't"; "you can't tell"; "can't tell me" |

N-Gram Features count the number of times each n-gram occurs in a Justice's questions to a particular side and serve as a numerical representation of the vocabulary and grammatical patterns in that Justice's questions.

N-grams have long been relied on to measure and analyze written texts—most traditionally to identify writing styles characteristic of particular authors or time periods. For example, using n-grams to measure the linguistic and metrical patterns of known works and time periods, scholars have been able to date various plays of Euripides for which external dating evidence is unavailable[28] and to authenticate *Double Falsehood*, published more than 100 years after Shakespeare's death, as a true work of the playwright.[29]

---

[27] Black et al., *supra* note 9, at 575.

In recent years, n-grams have been applied to an ever-growing, seemingly infinite variety of tasks, from identifying gender differences in tweeting style and detecting spam blogs to predicting a swine flu pandemic.[30] N-grams have also proven useful in studying various legal issues, for example, identifying the authors of unsigned judicial opinions[31] and studying the evolution of legal language over time.[32]

Given their success capturing and measuring lexical patterns across such a wide variety of contexts, n-grams should be a powerful aid to the study of oral argument—in particular how Justices' questioning styles vary when they are skeptical versus supportive of an advocate's position.

To collect the n-gram counts, the author created a Python script to count for each case the number of occurrences of every n-gram in each Justice's questions to each party and store the tallies in a database. Before the n-gram counts were collected, a separate script normalized[33] the text by converting all words to lowercase (so that, for example, n-grams will be treated as identical whether or not they start a sentence); removing punctuation and other nonalphanumeric characters; and removing so-called "stop words," such as articles, pronouns, and prepositions, which carry comparatively little meaning and generally decrease the effectiveness of n-gram feature analysis. The Justices' questions to advocates over the study period included 15,418,510 unique n-grams of one to five words in length. N-grams were limited to five words in length because n-grams longer than five words would recur infrequently from case to case and not contribute meaningfully to the analysis. Because separate n-gram counts were required for questions to each side, a total of 30,837,020 n-gram counts were tallied for each Justice in each case, representing the number of times a Justice spoke every possible n-gram to each side.

---

[28] *See generally* A. M. DALE, HELEN xxiv–xxviii (Oxford 1967) (explaining the chronology of Euripidean tragedies).

[29] *See* Ryan L. Boyd & James W. Pennebaker, *Did Shakespeare Write* Double Falsehood*? Identifying Individuals by Creating Psychological Signatures with Text Analysis*, 26 PSYCHOL. SCI. 570 (2015).

[30] *See* Zachary Miller et al., *Gender Prediction on Twitter Using Stream Algorithms with N-Gram Character Features*, 2 INT'L J. OF INTELLIGENCE SCI. 143 (2012); Pranam Kolari et al., *Detecting Spam Blogs: A Machine Learning Approach*, 21 NAT. CONF. ON ARTIFICIAL INTELLIGENCE (2006); Joshua Ritterman et al., *Using Prediction Markets and Twitter to Predict a Swine Flu Pandemic*, in Workshop on Mining Social Media (2009).

[31] *See* William Li et al., *Using Algorithmic Attribution Techniques to Determine Authorship in Unsigned Judicial Opinions*, 16 STAN. TECH. L. REV. 503 (2013).

[32] Daniel Martin Katz et al.*, Legal N-Grams? A Simple Approach to Track the 'Evolution' of Legal Language*, 24 JURIX: INT'L CONF. ON LEGAL KNOWLEDGE INFO. SYS (2011).

[33] Text normalization is a standard step in natural language processing tasks. Normalization ensures that similar phrases (for example, "I'd like to hear the answer" and "But I would like to hear your answer") are recognized as approximate equivalents despite slight differences in phrasing or differences in punctuation, capitalization, or tense.

### III. Summary of Data

Before describing the construction of predictive models, this Section pauses to review the extracted feature data at a high level, present various highlights, and gauge the features' likely predictive power.

#### A. Question Counts

A high-level review of the Question Count Features data collected for this study shows significant differences among the Justices in their questioning behavior at oral argument. The data also bolster the conclusion of past authors[34] who have observed a correlation between questioning intensity and the Justices' votes: The lengthier and more numerous a Justice's questions to a party, the less likely that party is to get the Justice's vote.

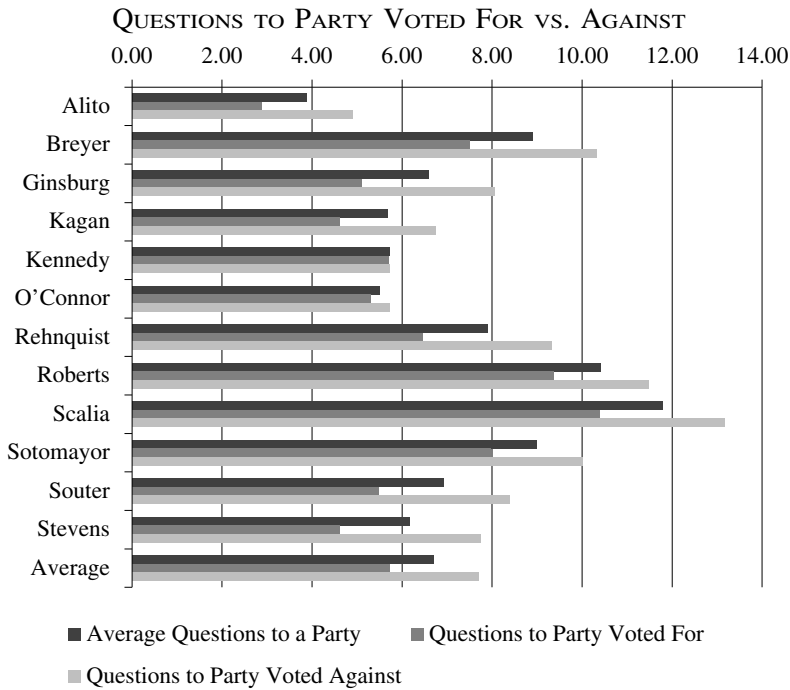

QUESTIONS TO PARTY VOTED FOR vs. AGAINST

■ Average Questions to a Party   ■ Questions to Party Voted For
■ Questions to Party Voted Against

---

[34] *See, e.g.,* Johnson et al., *supra* note 7; EPSTEIN ET AL., *supra* note 3; Black et al., *supra* note 9.

As shown in the table above,[35] the intensity of questioning varies considerably from Justice to Justice, with the least active questioner, Justice Alito, asking only about a third as many questions per side (3.89) as the most active questioner, Justice Scalia (11.78). Justices Breyer, Roberts, Scalia, and Sotomayor stand out as the most active questioners, asking between 8.98 and 11.78 questions per side; Justice Alito asks notably fewer questions than the other Justices, on average only 3.89 per side; and the remaining seven Justices fall somewhere around the median of 6.71 questions per side.

For almost all of the Justices, the number of questions they ask to each side correlates with their ultimate votes, with more intense questioning going to the eventual loser. This pattern supports the widely held view that a hot bench is a bad sign for advocates. True to their reputations as swing Justices, however, Justices Kennedy and O'Connor buck the trend, barely differing in their questioning between parties they ultimately vote for versus against.

|  | Mean Questions | | | Words Per Question | | |
| --- | --- | --- | --- | --- | --- | --- |
|  | Voted For | Voted Against | Difference | Voted For | Voted Against | Difference |
| Alito | 2.87 | 4.91 | 2.04 | 46.05 | 43.99 | -2.06 |
| Breyer | 7.49 | 10.31 | 2.82 | 49.70 | 53.82 | 4.12 |
| Ginsburg | 5.11 | 8.06 | 2.94 | 38.40 | 40.35 | 1.95 |
| Kagan | 4.61 | 6.75 | 2.14 | 49.80 | 53.85 | 4.05 |
| Kennedy | 5.71 | 5.72 | 0.02 | 29.02 | 30.87 | 1.85 |
| O'Connor | 5.30 | 5.71 | 0.41 | 19.77 | 22.09 | 2.32 |
| Rehnquist | 6.46 | 9.32 | 2.86 | 15.41 | 18.10 | 2.69 |
| Roberts | 9.36 | 11.47 | 2.10 | 24.76 | 27.94 | 3.18 |
| Scalia | 10.40 | 13.16 | 2.77 | 27.69 | 32.09 | 4.40 |
| Sotomayor | 7.96 | 10.00 | 2.04 | 31.92 | 33.50 | 1.58 |
| Souter | 5.47 | 8.39 | 2.92 | 38.89 | 43.95 | 5.06 |
| Stevens | 4.61 | 7.74 | 3.13 | 26.95 | 25.96 | -0.99 |
| Average | 5.72 | 7.70 | 1.98 | 33.41 | 36.29 | 2.88 |

Similar trends apply to the wordiness of the Justices' questions, shown in the table above alongside the question-count data. All except Justices Alito and Stevens ask shorter-than-average questions to parties for whom they ultimately vote and longer-than-average questions to the side they vote against. This disparity may be attributable to the different sorts of questions that Justices will employ depending on whether they are persuaded by or skeptical of an advocate's position: Complicated hy-pothetical questions designed to test or critique will be longer than quick "bolstering" questions designed to help an advocate recover from a difficult series of questions.

---

[35] Here, and throughout the Article, Justice Thomas is excluded from the analysis. Justice Thomas almost never participates in oral argument, preventing any meaningful analysis of his questioning behavior. *See* Adam Liptak, *A Thomas Milestone Likely to Pass Quietly*, N.Y. TIMES, Feb. 2, 2016, at A20 (noting Justice Thomas's ten-year silence).

Also notable is the trend toward longer questions among the more recently appointed members of the Court. Sometimes dubbed the "law professor" effect, the average number of words per question has steadily risen over the last 30 years,[36] beginning with the appointment of Justice Scalia in 1986 and rising with the additions of Justices Ginsburg, Breyer, and Kagan—all former law professors who pose more ponderous questions than their predecessors.[37]

## B. Question Chronology

The Question Chronology Features show that the Justices differ not only in the number of questions they ask at oral argument but also in the timing of those questions. Some jump right into questioning at the start of argument; others hold back and observe before asking questions; some ask one-off questions while others come back with multiple follow-up questions; and nearly all vary their pattern when questioning parties they ultimately vote for versus against.

The table below shows the percentage of arguments in which that Justice was the first to ask a question of an advocate. The Justices' questioning patterns vary significantly. As one would expect, some of the Court's most active questioners tend to lead off. This holds true for Justices Roberts, Scalia, and Sotomayor especially. But there are exceptions. One of the Court's most active questioners at oral argument, Justice Breyer, asks the first question in only about 5% of cases, well below the mean, matching his reputation as an especially deliberate questioner who "seem[s] to seek a dialogue with the advocate."[38] And Justice O'Connor, below average in question volume, is nonetheless the first to ask a question in more than 22% of cases.[39]

---

[36] *See* Johnson et al., *supra* note 7, at 251–53.

[37] *See* EPSTEIN ET AL., *supra* note 3, at 330–36 (noting similar results even without considering Justice Kagan, who has turned out to be an even more verbose questioner than Justice Scalia).

[38] WRIGHTSMAN, *supra* note 6, at 102.

[39] This surprising statistic comports with the comments of another author who, after observing a series of arguments in the 2002 term noted that at oral argument Justice O'Connor "was quite impatient with positions that she saw as patently ridiculous [and] had no reluctance in saying so. . . . [W]hen the petitioner's attorney was initially interrupted during his or her opening statement, it was most often by Justice O'Connor, Justice Scalia, or Justice Souter." WRIGHTSMAN, *supra* note 6, at 102.

|  | No. Cases 1st Question to Pet'r | No. Cases 1st Question to Resp't | Total Arguments | % Arguments Asked First Question |
|---|---|---|---|---|
| Alito | 32 | 55 | 1616 | 5.38% |
| Breyer | 31 | 79 | 2766 | 3.98% |
| Ginsburg | 261 | 145 | 2768 | 14.67% |
| Kagan | 21 | 31 | 896 | 5.80% |
| Kennedy | 147 | 105 | 2764 | 9.12% |
| O'Connor | 179 | 89 | 1186 | 22.60% |
| Rehnquist | 187 | 140 | 1114 | 29.35% |
| Roberts | 117 | 180 | 1662 | 17.87% |
| Scalia | 204 | 236 | 2706 | 16.26% |
| Sotomayor | 93 | 103 | 1054 | 18.60% |
| Souter | 48 | 61 | 1712 | 6.37% |
| Stevens | 39 | 84 | 1864 | 6.60% |

Whether a Justice is the first to pose a question also reveals a lot about that Justice's view of the party's case. As shown in the chart below, a Justice who initiates questioning of one side at oral argument will ultimately vote for that side in only about 40% of cases. The strength of the tendency varies between Justices, with a first question from normally slow-to-speak Justice Breyer being an especially bad sign for advocates.

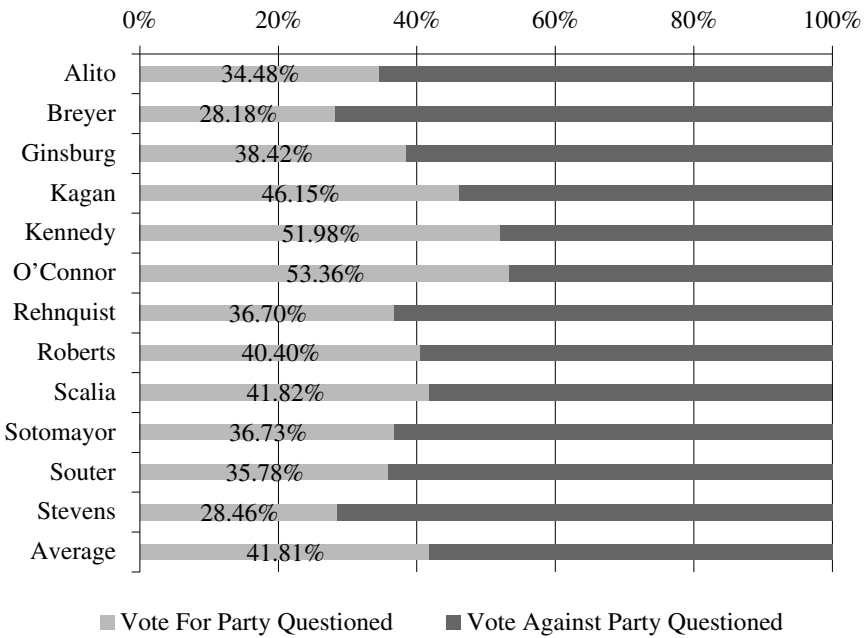

| Justice | Vote For Party Questioned |
|---|---|
| Alito | 34.48% |
| Breyer | 28.18% |
| Ginsburg | 38.42% |
| Kagan | 46.15% |
| Kennedy | 51.98% |
| O'Connor | 53.36% |
| Rehnquist | 36.70% |
| Roberts | 40.40% |
| Scalia | 41.82% |
| Sotomayor | 36.73% |
| Souter | 35.78% |
| Stevens | 28.46% |
| Average | 41.81% |

Even beyond the very first question of an oral argument, how long a Justice waits before asking her first question to a party correlates with how that Justice ultimately votes in the case. The chart below compares the average timing of each Justices' first question with the average timing of that Justices' first question to the party she ultimately votes for and against. All except Justices Kennedy and O'Connor wait on average deeper into oral argument before asking a question of the party they ultimately vote for.

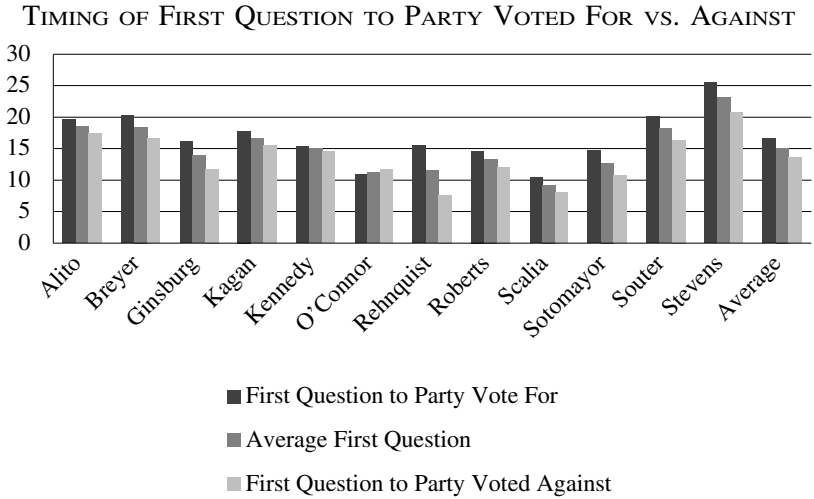

TIMING OF FIRST QUESTION TO PARTY VOTED FOR VS. AGAINST

■ First Question to Party Vote For
■ Average First Question
■ First Question to Party Voted Against

The Justices vary, too, in how many follow-up questions they tend to ask. Some tend to ask multiple one-off questions, while others engage in more sustained questioning. The table below shows the average number of consecutive questions each Justice asks to the party she ultimately votes for and against. Questions were deemed consecutive whenever, following an advocate's response, a Justice asked an additional question without another Justice intervening to pose her own question.

Mean Consecutive Questions

|  | Party Voted For | Party Voted Against | Difference |
| --- | --- | --- | --- |
| Alito | 1.94 | 2.22 | 0.27 |
| Breyer | 3.27 | 3.54 | 0.27 |
| Ginsburg | 1.72 | 2.18 | 0.45 |
| Kagan | 1.82 | 2.14 | 0.32 |
| Kennedy | 2.19 | 2.22 | 0.02 |
| O'Connor | 2.28 | 2.51 | 0.23 |
| Rehnquist | 1.49 | 1.91 | 0.43 |
| Roberts | 1.81 | 2.08 | 0.27 |
| Scalia | 2.28 | 2.63 | 0.35 |
| Sotomayor | 2.99 | 3.21 | 0.22 |
| Souter | 2.55 | 3.05 | 0.50 |
| Stevens | 2.30 | 2.84 | 0.54 |
| Average | 2.25 | 2.59 | 0.34 |

As it so happens, Justice Breyer, the wordiest questioner, is also the most likely to follow up with an additional question. On average, he asks more than three questions in a row before another Justice interjects. Given the volume of their questioning, Justices Scalia and Sotomayor naturally also top the list of those most likely to ask a follow-up question. Others generally follow the same trend, with the most voluminous questioners also being the most likely to engage in sustained, multiquestion dialogues with the advocates. The exception is Chief Justice Roberts, who ranks among the Court's most frequent questioners, but whose questions are shorter than his colleagues and less likely to be supplemented with a follow-up question.

The intensity of nearly all of the Justices' questioning differs between parties they ultimately vote for versus against, with the more sustained questioning going to the ultimate loser. On average, each Justice asks 0.34 more questions per series of consecutive questions to the party she votes for. As with the other metrics, the exception is the Court's swing vote, Justice Kennedy. He asks the same number of follow-up questions on average to both sides. Interestingly, though, the Court's other traditional swing vote, Justice O'Connor, doesn't hold her cards quite so close. Justice O'Connor asks approximately the same total number of questions per side, like Justice Kennedy. But unlike Justice Kennedy and like the rest of the Court, Justice O'Connor tends to engage in longer question exchanges with the party she ultimately votes against, with average exchanges of 2.51 versus 2.28 consecutive questions.

C. *Question Sentiment*

The Question Sentiment data collected show differences between the Justices in the tone of their questioning at oral argument. The data also confirm the work of past authors who have observed differences in the emotional content of the Justices' questions to parties they ultimately vote for versus against.[40]

The table below shows the average sentiment of each Justice's questions, as measured by the Stanford CoreNLP sentiment annotator, to all parties collectively, to the party the Justice ultimately voted for, and to the party ultimately voted against. The sentiment scale ranges from one to five, with one representing a very negative statement and five representing a very positive one. The average sentiment of oral argument questions across all Justices falls between negative and neutral, at 2.48. Chief Justices Rehnquist and Roberts ask the friendliest questions on av-erage, Justices Kagan and Alito the least friendly, and the rest of the Justices fall somewhere around the average.

---

[40] *See* Black et al., *supra* note 9, at 576–77 (finding difference in percent pleasant versus unpliant words directed to each side to be correlated with both case outcomes and individual justice votes); Dietrich et al., *supra* note 25, at 4–7 (emotional arousal of Justices measured by vocal pitch of audio recordings correlated with Justices' votes).

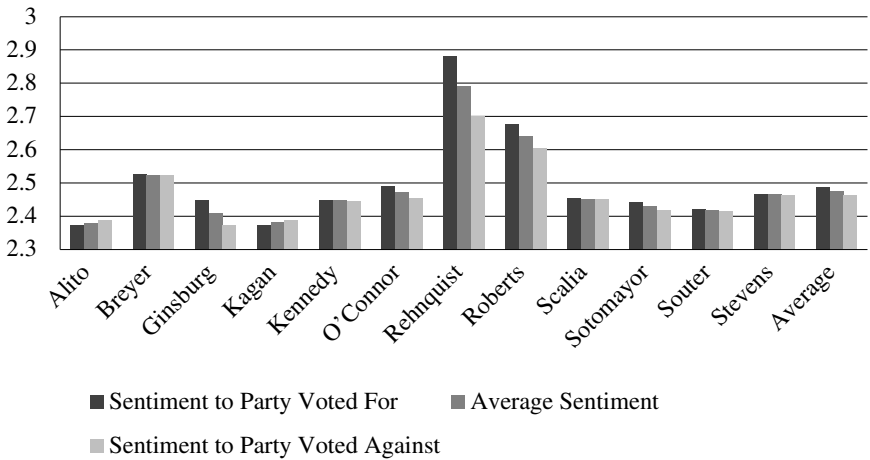

SENTIMENT OF QUESTIONS TO PARTY VOTED FOR VS. AGAINST

Many, but not all, of the Justices show differences in question sentiment between the party they ultimately vote for versus against. Justices Rehnquist and Ginsburg show the greatest changes in tone based on which side they are questioning. Justices Breyer, Scalia, Stevens, and Kennedy show almost no difference. And Justices Alito and Kagan ask *friendlier* questions on average to the party they ultimately vote against.

D.  N-Grams

Not only do the data show differences in questioning patterns between Justices—for example, more sustained, or lengthier questions by some Justices than others—the data also show differences in the questions themselves. Each Justice has her own unique questioning style. And the same Justice will ask different sorts of questions and differently phrased questions to advocates depending on whether the Justice is skeptical of or favors the advocate's position.

The table below gives an overview of the n-gram data collected for this study.

|           | Total Distinct N-Grams Spoken |
|-----------|------------------------------:|
| Alito     | 617,427                       |
| Breyer    | 2,470,241                     |
| Ginsburg  | 1,549,705                     |
| Kagan     | 557,235                       |
| Kennedy   | 974,395                       |
| O'Connor  | 277,751                       |
| Rehnquist | 326,893                       |
| Roberts   | 1,074,699                     |
| Scalia    | 1,876,251                     |
| Sotomayor | 649,481                       |
| Souter    | 1,115,793                     |
| Stevens   | 638,937                       |

Over the study period, 1998–2015, each Justice's questions included between 500,000 and 2.5 million distinct n-gram phrases of between one and five words in length. The difference in total distinct n-grams is largely attributable to two factors: (1) the number of oral arguments the Justice participated in during the study period; and (2) the number and length of questions the Justice asked during those arguments.

Justices Breyer and Scalia top the list as frequent and verbose questioners who served during the entire study period. By contrast, Justices O'Connor and Rehnquist, whose tenures ended early in the study period, and President Obama's appointees, Justices Sotomayor and Kagan, trail behind their longer-serving colleagues. Predictive models are likely to be more accurate for those Justices who speak most often at oral argument and who participated in the greatest number of oral arguments during the study period.

## IV. MODEL TESTING AND RESULTS

A high-level review of the data suggests that each category of features—Question Count Features, Question Chronology Features, Question Sentiment Features, and N-Gram Features—provides at least some measure of how a Justice perceives an advocate's position. Almost all of the Justices ask more and longer questions of the party they ultimately vote against; the questions are less friendly in tone; and question exchanges include more follow-up questions.

Having explored the intuitive relevance of these features, we will turn to the construction of Justice voting models to test the features' predictive power. Rather than create one single Justice voting model, separate models were created for each Justice to account for each Justice's individual questioning style. Feature matrices[41] were created for each Justice, with each row in each matrix representing the Justice's questioning to a party at oral argument, one column indicating whether the Justice voted for or against the party, and more than 30 million additional columns representing the extracted features: how many questions the Justice asked; how many consecutive questions on average; the questions' average sentiment; and, importantly, columns for each of the millions of unique n-gram phrases, a column indicating how many times the Justice spoke each phrase to the party (for most phrases, typically 0), and another set of columns indicating how many times the Justice spoke each phrase to the party's opponent.

These feature matrices were used to construct Support Vector Machine (SVM) models for each Justice in Python using the LinearSVC class of scikit-learn 0.18. Support Vector Machines were chosen rather than Random Forests, Extremely Randomized Trees, AdaBoost Classifiers, and other popular alternative classifiers because SVMs are especially effective in high dimensional feature spaces and have performed well in analogous contexts such as n-gram-based text classification and authorship attribution.[42]

A.  *Vote Prediction Accuracy*

The predictive power of the Justice SVM models was tested using a standard k-fold cross-validation procedure with ten folds. Each feature matrix was repeatedly divided into ten samples, nine of which were used to train an SVM model, and the tenth of which, the holdout sample, was not used in training and was set aside to estimate the model's accuracy. This procedure simulates the performance of the models in predicting Justice votes in future, argued-but-not-decided cases based on historical oral argument data and outcomes.

---

[41] A feature matrix is a two-dimensional representation of the input data to be used for training and testing a predictive model. Here, each row in the matrix represents a Justice's interaction with one side during oral argument in a particular case, and each column in that row constitutes a single data point describing one feature of the Justice's questioning behavior, for example, how many questions the Justice asked that party, or how many times he or she spoke the words "well, I think."

[42] Authorship attribution is the process of determining the likely author of a document of unknown authorship based on its similarity to other documents with known authors. Text classification is the process of assigning a document to one of multiple potential categories based on the words it contains. E-mail spam filtering, for example, is a familiar text-classification task, which requires an e-mail to be categorized either as "spam" or "not spam" based on its content. N-gram–based solutions to such problems typically involve high dimensional feature spaces (large numbers of features) because of the breadth of human vocabulary. Here, Justice vote prediction can be conceptualized as a text-classification problem: A Justice's oral argument questions are categorized as the questions of a Justice who votes for or against the party being questioned.

The table below shows the models' performance predicting individual Justice votes in all cases in which the Justice participated during the study period. For cases beginning with the Court's 1998 term and running through the 2015 term, the Justice SVM models correctly predicted individual Justice votes in 73% of the cases.

| | Total Arguments | Correctly Predicted | Baseline (Reversal %) | Accuracy |
|---|---|---|---|---|
| Alito | 808 | 518 | 0.592 | 0.641 |
| Breyer | 1383 | 1034 | 0.617 | 0.748 |
| Ginsburg | 1384 | 1067 | 0.577 | 0.771 |
| Kagan | 448 | 271 | 0.549 | 0.605 |
| Kennedy | 1382 | 1036 | 0.648 | 0.749 |
| O'Connor | 593 | 506 | 0.641 | 0.854 |
| Rehnquist | 557 | 484 | 0.636 | 0.868 |
| Roberts | 831 | 497 | 0.635 | 0.598 |
| Scalia | 1353 | 1025 | 0.615 | 0.757 |
| Sotomayor | 527 | 276 | 0.581 | 0.523 |
| Souter | 856 | 669 | 0.609 | 0.782 |
| Stevens | 932 | 705 | 0.565 | 0.756 |
| All | 11054 | 8088 | 0.607 | 0.732 |

Predictive accuracy varies significantly between Justices, from a low of 52% for Justice Sotomayor, to a high of 87% for Justice Rehnquist. This disparity is attributable in part to the different quantity of training data available depending on the dates of the Justice's service. Because a separate model was created for each Justice, in general, the more historical argument data is available for that Justice, the more accurate the resultant model.

The models' performance compares favorably to recent work by others who have developed models that predict Justice votes with anywhere from 57% to about 73% accuracy, depending on the particular input data, model, and period studied.[43] The models also outperform a baseline petitioner-always-wins model, a measure that is commonly used to benchmark predictive models because of the Supreme Court's relatively steady 60% reversal rate.[44]

---

[43] *See* Dietrich et al., *supra* note 25, at 4–6 (56% vote prediction accuracy using Justice voice pitch); Chen et al., *supra* note 10, at 7–8 (67% vote prediction accuracy using audio and facial attribute features of advocates); Daniel Martin Katz et al., *A General Approach for Predicting the Behavior of the Supreme Court of the United States*, PLOS ONE at 8 (2017) (72% accuracy using data from the Washington University School of Law's Supreme Court Database, including information regarding the issue on appeal and the decision of the lower court); Black et al., *supra* note 9, at 576–77 (73% accuracy using question counts, question pleasantness, and case-specific ideological measures).

[44] Roy E. Hofer, *Supreme Court Reversal Rates: Evaluating the Federal Courts of Appeals*, 2 LANDSLIDE, January/February 2010 at 1, https://www.americanbar.org/content/dam/

The models also offer key advantages over prior Justice-voting models. First, the models are tailored to individual Justices. The models take into account each Justice's particular questioning patterns and will thus be useful for examining not only question behavior generally but also how the Justices' particular oral argument patterns differ from one another.

Second, the models rely entirely on data internal to the oral argument itself—question counts, sentiment, n-grams etc.—and not on external factors such as Justice ideology, ideology of the lower-court decision, identity of the parties, or the particular legal issue on appeal. The models are thus especially useful for examining the goings on at oral argument rather than voting behavior more generally. Moreover, the models' extremely high predictive accuracy even without taking into account features such as justice and case ideology, suggests that oral argument questioning reveals information not fully reflected by standard ideological measures. N-grams and other oral argument data could likely be combined with external measures such as Justice and case ideology to produce more powerful general predictive models.

## B. *Performance of Individual Feature Categories*

The previous section described the predictive power of Justice SVMs trained using all of the extracted features: Question Counts, Question Chronology, Question Sentiment, and N-grams. To gauge the incremental contribution of each feature category, a series of models were also created for each Justice using various subsets of the available features. The table below shows the performance of Justice SVM models constructed using single categories of features.

|  | Question Counts | Question Chronology | Question Sentiment | N-Grams | Baseline (Reversal %) |
|---|---|---|---|---|---|
| Alito | 0.599 | 0.626 | 0.616 | 0.649 | 0.592 |
| Breyer | 0.602 | 0.628 | 0.630 | 0.639 | 0.617 |
| Ginsburg | 0.614 | 0.650 | 0.582 | 0.667 | 0.577 |
| Kagan | 0.556 | 0.591 | 0.604 | 0.639 | 0.549 |
| Kennedy | 0.641 | 0.651 | 0.652 | 0.632 | 0.648 |
| O'Connor | 0.667 | 0.666 | 0.666 | 0.683 | 0.641 |
| Rehnquist | 0.712 | 0.677 | 0.652 | 0.720 | 0.636 |
| Roberts | 0.595 | 0.648 | 0.648 | 0.618 | 0.635 |
| Scalia | 0.569 | 0.632 | 0.632 | 0.650 | 0.615 |
| Sotomayor | 0.531 | 0.591 | 0.602 | 0.628 | 0.581 |
| Souter | 0.598 | 0.653 | 0.666 | 0.635 | 0.609 |
| Stevens | 0.606 | 0.584 | 0.563 | 0.594 | 0.565 |
| All | 0.577 | 0.623 | 0.622 | 0.644 | 0.607 |

aba/migrated/intelprop/magazine/LandslideJan2010_Hofer.authcheckdam.pdf (finding a median reversal rate of 68%).

In contrast with the all-features models discussed previously, no model based on a single feature category significantly outperformed its baseline. That is especially surprising in the case of the N-Grams models. The N-Grams models, recall, include millions of features that as a whole constitute the entire corpus of each Justice's oral argument questions in each case—counts of every word and phrase spoken. It might have been expected that the n-gram features would render all other feature categories superfluous and that the N-Grams-only models would perform equally with the all-feature models. What use are question counts or numerical measures of question sentiment, after all, if one has already recorded counts of every word spoken in every question? But that is not the case. The N-Grams-only models performed only slightly better than other single-feature-category models.

As the table below shows, each category of features generally adds some incremental predictive accuracy beyond the n-grams alone.

|           | N-Grams | N-Grams, Counts | N-Grams, Counts, Chronology | N-Grams, Counts, Chronology, Sentiment |
|-----------|---------|-----------------|-----------------------------|----------------------------------------|
| Alito     | 0.649   | 0.640           | 0.644                       | 0.641                                  |
| Breyer    | 0.639   | 0.662           | 0.702                       | 0.748                                  |
| Ginsburg  | 0.667   | 0.691           | 0.734                       | 0.771                                  |
| Kagan     | 0.639   | 0.612           | 0.629                       | 0.605                                  |
| Kennedy   | 0.632   | 0.661           | 0.697                       | 0.749                                  |
| O'Connor  | 0.683   | 0.757           | 0.810                       | 0.854                                  |
| Rehnquist | 0.720   | 0.750           | 0.826                       | 0.868                                  |
| Roberts   | 0.618   | 0.588           | 0.588                       | 0.598                                  |
| Scalia    | 0.650   | 0.681           | 0.711                       | 0.757                                  |
| Sotomayor | 0.628   | 0.523           | 0.553                       | 0.523                                  |
| Souter    | 0.635   | 0.688           | 0.733                       | 0.782                                  |
| Stevens   | 0.594   | 0.648           | 0.695                       | 0.756                                  |
| All       | 0.644   | 0.662           | 0.697                       | 0.732                                  |

C.   *Weighting Features Within the Models*

Although each feature category improves overall predictive accuracy, not all features contribute equally. The table below shows the weight given by the models to each category of features collectively.

|  | Question Count Features | Question Chronology Features | Question Sentiment Features | N-Gram Features | Party |
|---|---|---|---|---|---|
| Alito | 0.174 | 0.037 | 0.049 | 0.571 | 0.169 |
| Breyer | 0.126 | 0.017 | 0.063 | 0.674 | 0.120 |
| Ginsburg | 0.173 | 0.061 | 0.043 | 0.581 | 0.142 |
| Kagan | 0.169 | 0.047 | 0.011 | 0.643 | 0.129 |
| Kennedy | 0.059 | 0.033 | 0.111 | 0.672 | 0.126 |
| O'Connor | 0.148 | 0.060 | 0.119 | 0.495 | 0.178 |
| Rehnquist | 0.424 | 0.068 | 0.185 | 0.245 | 0.078 |
| Roberts | 0.274 | 0.037 | 0.010 | 0.624 | 0.055 |
| Scalia | 0.117 | 0.061 | 0.020 | 0.686 | 0.116 |
| Sotomayor | 0.222 | 0.093 | 0.065 | 0.612 | 0.008 |
| Souter | 0.053 | 0.031 | 0.239 | 0.575 | 0.101 |
| Stevens | 0.153 | 0.072 | 0.017 | 0.614 | 0.144 |
| All | 0.156 | 0.048 | 0.073 | 0.604 | 0.118 |

The importance of each feature category varies from Justice to Justice in some expected ways. For example, Question Count Features constitute a modest 5.9% of Justice Kennedy's model, in contrast with a striking 42.4% of Justice Rehnquist's model. This accords with differences in the two Justice's questioning behaviors. During the study period, Justice Rehnquist posed an average of 2.86 more questions to advocates he voted against, whereas Justice Kennedy posed a near equal number of questions to both sides. The feature category is much more informative for Justice Rehnquist than for Justice Kennedy.

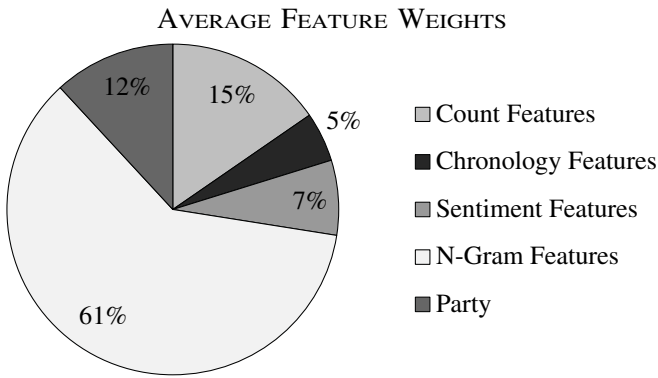

AVERAGE FEATURE WEIGHTS

The chart above shows average weight given to each feature category across all models. The N-Gram Features account for the majority of models' predictive power, but each of the other features also plays a sizeable role. It is surprising that the Chronology Features account for only a modest 5% of the models, given that models constructed from Chronology Features alone outperformed models constructed from Count or Sentiment Features alone. The relatively small contribution of the Sentiment Features is also surprising given the sophistication of the Stan-ford CoreNLP sentiment annotator and past work[45] that has demonstrated the usefulness of even rudimentary measures of sentiment. The low average weight of Sentiment Features may be explained by a high correlation between a question's sentiment and the n-grams it contains.

## V. THE JUSTICES' QUESTIONING STYLES

Beyond their impressive predictive power, the Justice SVM models, particularly their N-Gram Features, also offer a new window into oral argument. Whereas past study of oral argument has relied on basic aggregate measures (e.g., how many questions the justices ask)[46] or human observers' intuitive evaluations,[47] the N-Gram Features offer a quantitative tool with which to examine the content of the Justice's questions. By identifying and examining key n-grams that are highly predictive of each Justice's eventual vote, we gain insights into some of the ways the Justices make use of oral argument.

Before moving on to review particular n-grams, a few general observations:

> Without exception, the Justices reveal their leanings more clearly through negative things they say to the parties they ultimately vote against than positive things they say to parties they vote for. For every one of the Justices, "negative" n-grams were much more powerful predictors of the Justice's ultimate vote than were "positive" n-grams.

The most highly predictive n-grams are often topic-invariant phrases that reflect the Justice's decision-making processes (e.g. expressing skepticism or introducing a hypothetical question) rather than the subject matter of the particular case.

Many of the highly predictive n-grams correlate with the Justices' votes because they reveal the Justices' attitudes toward a party directly (e.g., direct criticism). But others are correlated with the Justices' votes for reasons less directly related to oral argument itself, because, for example, they include the identity of the particular advocate (e.g., Paul Clement) or indicate whether a party is the petitioner or respondent in the case.

The table below presents a few (of thousands) of the most highly predictive positive and negative n-grams for each of the Justices. Positive n-grams are those correlated with a vote for the party to whom they are spoken, whereas negative n-grams are those correlated with a vote against the party to whom they are spoken.

---

[45] *See, e.g.,* Black et al., *supra* note 9.
[46] *See, e.g.,* Johnson et al., *supra* note 7; EPSTEIN ET AL., *supra* note 3; Black et al., *supra* note 9; Chen et al., *supra* note 10.
[47] *See, e.g.*, MALPHURS, *supra* note 4, at 60–101; WRIGHTSMAN *supra* note 6.

| Justice | | Highly Predictive N-Grams |
|---|---|---|
| Alito | + | "adopt", "interpret", "near", "work", "correct", "everyth", "petition", "make argument", "take case", "rule would appli" |
| | – | "sinc", "pass", "object", "trouble", "undermin", "thousand", "make differ", "question whether", "say district court" |
| Breyer | + | "respons", "little", "ordinary", "self", "neglig", "central", "could say", "want say", "far concern", "suppose accept argument" |
| | – | "rather", "grant", "seem", "constitute", "new", "foresee", "well even", "justice laughter", "decide whether", "say want know" |
| Ginsburg | + | "guilt", "came", "comparison", "mani peopl", "good deal", "take posit", "anyth els", "say go give", "justic kennedi question" |
| | – | "result", "congress", "compani", "would say", "congress said", "mean could", "chang law", "know mean", "would clarify one" |
| Kagan | + | "equival", "compani", "let", "argument", "true", "professor", "wonder whether", "court say", "justice breyer suggest" |
| | – | "thought", "well", "anyth", "question whether", "think congress", "say someth", "one would", "back justice scalia" |
| Kennedy | + | "charg", "context", "figure", "court get", "well problem", "say concern", "justic breyer", "want write opinion" |
| | – | "firearm", "baseless", "two case", "well govern", "think would", "say noth", "cours know answer", "would case differ" |
| O'Connor | + | "prison", "today", "interpret", "well guess", "take posit", "add mix", "let ask someth", "well say take", "let ask suppose" |
| | – | "depart", "overrul", "read", "case deal", "mean think", "well presume", "well statut", "well would think", "let ask anoth" |
| Rehnquist | + | "schlick", "alabama", "dreeben", "correct pronunci", "argument first morn number", "bright line rule", "said moment ago" |
| | – | "pardon", "hope", "argument", "well say", "say reject", "thank waxman", "crimin case", "well privat", "make differ whether" |
| Roberts | + | "shot", "fleme", "typic", "friend", "regulatori", "argument kent", "argument general", "go say well", "district court say" |
| | – | "see", "tribe", "fine", "mention", "would seem", "class action", "sorri follow", "blah blah", "justice stephen hypothet" |
| Scalia | + | "yeah", "presum", "well enough", "issu case", "well use", "use term", "rule mean", "statut would", "get rid", "would say one" |
| | – | "interim", "pretti", "potenti", "chang mind", "know whether", "logic conclus", "would would would", "justic ginsburg point" |
| Sotomayor | + | "okay", "kind", "move", "let assum", "see differ", "back justice", "right would", "give defer", "question justice", "go back justice" |
| | – | "congress", "never", "show", "congress intent", "chief justic", "posit take", "draw line", "could chang", "answer justic breyer" |
| Souter | + | "treat", "cover", "think", "court make", "go say", "understand argument", "know whether would", "justice kennedi question" |
| | – | "right", "well", "meant", "mean think", "govern say", "would conced", "justice breyer question", "theori behind may" |
| Stevens | + | "way", "contend", "hypothet", "ask question", "would appli", "may ask point", "ask question sort", "let clarify one thing" |
| | – | "explain", "erron", "true", "case kind", "cite case", "fact state", "understand correct", "may may interrupt", "one last question" |

As Lawrence Wrightsman and others have observed,[48] the Justices' questions at oral argument tend to fall into a few overlapping categories including (1) background questions seeking information missing or not highlighted in the briefing; (2) questions seeking to clarify an advocate's position or the scope of the rule being advocated; (3) questions about the limits of that rule or its implication for future cases, often framed as hypothetical questions; (4) questions offering support for an advocate's position; (5) questions criticizing an advocate's position; (6) questions designed primarily to communicate with other Justices; and (7) questions designed to interject humor into the argument and relieve tension.

The highly predictive n-grams show the Justices offering all of these types of questions at key moments during oral argument, each in her own unique way. A few examples are discussed in more detail below.

A.  *Information Seeking and Position Testing*

Oral argument gives the Justices an important opportunity to clarify the arguments in parties' briefs and to test the strength and scope of the positions they are advocating.[49] The highly predictive n-grams reveal the Justices regularly doing just that, each in her own way.

With one sort of question, the Justices convey genuine uncertainty about the direction the case should go and seek the advocate's assistance to clarify their views. Justice Kagan, for example, frequently frames her questions by stating "let me make sure I understand," summarizing her understanding of the advocate's position, and asking for confirmation or further explanation.[50] The phrase is a good sign for the advocate to whom it is asked, as it is closely correlated with Justice Kagan's vote for that party.

Similar uncertainty appears to underlie even some of the Justices' more aggressive-sounding questions. For example, Justice Kennedy commonly asks, "do you want us to write an opinion saying [some ex-treme view]"?[51] From Justice Kennedy, this sort of question conveys skepticism, but invites the advocate to describe a more reasonable alternative. Though threatening, the question appears to be a true invitation to dialogue and is correlated with Justice Kennedy's ultimate vote for the party.

---

[48] *See* WRIGHTSMAN, *supra* note 6, at 69–81 (presenting the categories above as one possible classification system and discussing alternative question-classification schemes); *see also* MALPHURS, *supra* note 4, at 86–92 (identifying fifteen purposes that the Justices accomplish through their questions at oral argument).

[49] *See* Mark Hummels, *Distributing Draft Decisions Before Oral Argument on Appeal: Should The Court Tip Its Tentative Hand? The Case for Dissemination*, 46 ARIZ. L. J. 325 (2004).

[50] *See, e.g.*, Transcript of Oral Argument at 45, Dep't of Homeland Sec. v. MacLean, 135 S. Ct. 913 (2015) (No. 13-894) ("Mr. Katyal, can we go back to your legal argument, and let me make sure I understand it."); Transcript of Oral Argument at 9–10, Coleman v. Court of Appeals, 566 U.S. 30 (2012) (No. 10-1016) ("So you are saying—let me just make sure I understand. You are saying that the—that Congress is thinking that an employer actually does think that women take more sick leave because women get pregnant.").

The import of a question is not always clear on its face. Even extremely similar questions can carry very different implications for the parties. For example, for Justice Ginsburg, the five-gram "would you clarify one thing" is one of the most powerful negative predictors among n-grams three words or longer. She frequently asks an advocate "would you clarify one thing?" in order to ensure her understanding of that party's position before ultimately rejecting it.[52] However, she also employs the almost identical phrase, "may I ask you to clarify?"[53] when asking for additional information from parties she ultimately sides with. The slight difference in phrasing is almost certainly subconscious, varying based on whether she expects to be persuaded or dissuaded by the advocate's response.

Of course, in many instances the Justices' perspectives are already nearly solidified. In such cases, the Justices come to argument armed with especially difficult questions, which ironically, they often introduce by first requesting permission. Justice Alito, for example, commonly begins by asking, "could I ask you this question?"[54] And Justice Kennedy similarly begins "let me ask you this question."[55] Though politely phrased, such questions imply confusion (at best) or doubt (more likely) regarding the advocate's position and are correlated with votes against the party to whom they are asked. That is especially so for the four-gram "let let me ask," which often appears in Justice Kennedy's questions when he stutters slightly while interrupting an advocate midsentence.[56] These sorts of questions help the Justices confirm their views while allowing the advocate a chance to persuade them otherwise.

---

[51] *See, e.g.*, Transcript of Oral Argument at 20, Air Wis. Airlines Corp. v. Hoeper, 134 S. Ct. 852 (2014) (No. 12-315) ("So—so you want us to write an opinion to say that the—that the statute here is to be interpreted differently than if it were a new York Times and Sullivan case or Masson-New Yorker case?"); Transcript of Oral Argument at 9, United States v. Grubbs, 547 U.S. 90 (2006) (No. 04-1414) ("I mean, if you want us to write the opinion with this qualification in it, it seems to me that you're making a big change in the way search warrants are used.").

[52] *See, e.g.*, Transcript of Oral Argument at 7, Carr v. United States, 560 U.S. 438 (2010) (No. 08-1301) ("[W]ould you clarify one thing? You're not questioning the Attorney General's determination that the underlying sex offense can have occurred pre-SORNA?"); Transcript of Oral Argument, Carmell v. Texas, 529 U.S. 513 (2000) (No. 98-7540) ("Mr. Bernstein, would you clarify one thing? You said something about 18 was the dividing line, but this child was . . . wasn't she 14?").

[53] *See, e.g.*, Transcript of Oral Argument at 20, Clark v. Arizona, 548 U.S. 735 (2006) (No. 05-5966) ("Mr. Goldberg, may I ask you to clarify one thing about your argument?"); Transcript of Oral Argument at 57, Morse v. Frederick, 551 U.S. 393 (2007) (No. 06-278) ("May I ask to you [*sic*] clarify one thing[?]").

[54] *See, e.g.,* Transcript of Oral Argument at 16, Loughrin v. United States, 134 S. Ct. 2384 (2014) (No. 13-316) ("Well, could I ask you this question? Suppose the defendant testifies and suppose the jury believes the defendant. The defendant testifies as follows . . . ."); Transcript of Oral Argument at 54, Dorsey v. United States, 567 U.S. 260 (2012) (No. 11-5683) ("Well, along those lines, could I—could I ask you this question, which is intended to explore the—the issue whether the argument about bringing the guidelines into consistency with applicable law doesn't assume the—the answer that is—that one attempts to get from it.").

B.  Unique Expressions of Skepticism

When observing oral argument, one can hardly help but notice the steady stream of critical and skeptical questions. Such questions serve to probe an advocate's position and test its strength or expose its weakness to a Justice's colleagues.[57] Without exception, n-grams introducing criticism are highly predictive in every one of the Justice models. The n-grams also show that each Justice has her own way of expressing that criticism.

Chief Justice Roberts for example, often notes that an advocate has proposed an "odd way"[58] of reading a statute or regulation or that it "would seem"[59] to him that the law should be understood another way. Justice Sotomayor expresses the same idea slightly differently, commonly noting that she "always thought"[60] the law to be something other

---

[55] *See, e.g.,* Transcript of Oral Argument at 37, Zedner v. United States, 547 U.S. 489 (2006) (No. 05-5992) ("Well, it seems to me the Government is equally remiss for not pointing out the obligations of the court under the act. But let me ask you this."); Transcript of Oral Argument at 24, Flores-Villar v. United States, 564 U.S. 210 (2011) (No. 09-5801) ("Let me just ask you this as an analytic matter, or as matter of logical priorities.").

[56] *See, e.g.*, Transcript of Oral Argument at 37, United States v. Dominguez Benitez, 542 U.S. 74 (2004) (No. 03-167) ("Let me . . . let me ask you this question. You argue for a subjective test in a context in which the defendant can't take the stand to say what his understanding was. That doesn't make a lot of sense to me."); Transcript of Oral Argument at 16, Ark. Dep't of Health and Human Servs. v. Ahlborn, 547 U.S. 268 (2006) (No. 04-1506) ("Let me . . . let me ask you this question so far as the rights of the assignee and assignor.").

[57] *See* Stephen M. Shapiro, *Oral Argument in the Supreme Court of the United States*, 33 CATH. U. L. REV. 529, 544, 547 (1984).

[58] *See, e.g.*, Transcript of Oral Argument at 4, D.C. v. Heller, 554 U.S. 570 (2008) (No. 07-290) ("Mr. Dellinger, it's certainly an odd way in the Second Amendment to phrase the operative provision."); Transcript of Oral Argument at 42, Kloeckner v. Solis, 568 U.S. 41 (2012) (No. 11-184) ("Well, yes, that's where the phrase comes in, but it does seem an odd way to establish that that is the critical element . . . .").

[59] Transcript of Oral Argument at 8, Heien v. N.C., 135 S. Ct. 530 (2014) (No. 13-604) ("It would seem to me that there's a stronger argument for taking the reasonableness of the officer's actions into account when you're talking about mistake of law . . . ."); Transcript of Oral Argument at 42, Horne v. Dep't of Agric., 135 S. Ct. 2419 (2015) (No. 14-275) ("The answer—I mean, if the answer is always you can do something else, it would seem we should—we'll never have these kinds of cases.").

[60] *See, e.g.*, Transcript of Oral Argument at 12, Universal Health Servs. v. United States *ex rel.* Escobar, 136 S. Ct. 1989 (2016) (No. 15-7) ("I'm sorry. I'm totally confused. I always

than what the advocate is suggesting or that she is "a little bit confused"[61] by the party's argument.

Many of the Justices' most negative n-grams are those that introduce pointed questions or, even worse, redirect an advocate back to a prior question (perhaps one that she was hoping to avoid). Each Justice has her own way of doing so. Justice Alito, for example, frequently tells advocates that he "would appreciate"[62] if counsel could explain something to him, while Justice Sotomayor, regularly asks counsel to "tell me what the difference is"[63] between two cases or factual scenarios. Several of the most negative n-grams appear when one of the Justices feels that an advocate has not satisfactorily answered a prior question and presses for a further answer. Justice Breyer, especially, often notes that he "would like to hear the answer"[64] to a past question.

The highly predictive n-grams also put on display each of the Justices' verbal tics. As just one of many examples, consider Justices Roberts. When presenting a line of reasoning he is skeptical of, Justice Roberts will often substitute "blah blah blah" for one or more steps in the argument:

> CHIEF JUSTICE ROBERTS: It's not a presentment requirement. That's in (a)(1). It's that the claim be paid, the fault claim, be paid by the

---

thought that when you asked for payment, you're making a promise: I did what I agreed to do. Pay me, please."); Transcript of Oral Argument at 31–32, Dorsey v. United States, 567 U.S. 260 (2012) (No. 11-5683) ("I always thought that when discrimination was at issue, that we should do as speedy a remedy as we could, because it is one of the most fundamental tenets of our Constitution . . . that our laws should be . . . enforced in a race-neutral way.").

[61] *See, e.g.*, Transcript of Oral Argument at 44, Montanile v. Bd. of Trs. of the Nat'l Elevator Indus. Health Benefit Plan, 136 S. Ct. 651 (2016) (No. 14-723) ("I'm still a little bit confused by all of this. In my mind, you get money when somebody gives it to you."); Transcript of Oral Argument at 27, T-Mobile S., LLC v. City of Roswell, 135 S. Ct. 808 (2015) (No. 13-975) ("I'm—I'm a little bit confused by that because and this is what troubles me the most.").

[62] *See, e.g.*, Transcript of Oral Argument at 29, Ariz. Free Enter. Club's Freedom Club PAC v. Bennett, 564 U.S. 721 (2011) (No. 10-238) ("Well, there are States that have public funding without having a matching fund provision. I would appreciate it if you would compare these two regimes."); Transcript of Oral Argument at 3, Carr v. United States, 560 U.S. 438 (2010) (No. 08-1301) ("Mr. Rothfeld, I wondered if I could ask you about three interrelated points concerning your textual argument. And if I could just lay those on the table and get your reaction to them, I would appreciate it.").

[63] *See, e.g.*, Transcript of Oral Argument at 9, United States v. Stevens, 559 U.S. 460 (2010) (No. 08-769) ("Could you—could you tell me what the difference is between these video [*sic*] and David Roma's documentary on pit bulls?"); Transcript of Oral Argument at 10, Skinner v. Switzer, 562 U.S. 521 (2011) (No. 09-9000) ("Could you tell me how that's different than what Alaska did in the Osborne case that we upheld their procedure?").

[64] *See, e.g.*, Transcript of Oral Argument at 13–14, Messerschmidt v. Millender, 565 U.S. 535 (2012) (No. 10-704) ("Now that was—the question, I think roughly, that you were being asked, and I would like to hear the answer."); Transcript of Oral Argument at 15, Roberts v. Sea-Land Servs., Inc., 566 U.S. 93 (2012) (No. 10-1399) ("Would you then go back—I did have the same question Justice Alito asked, and I'd like to hear the answer.").

> government. And what you're saying is when the
> government pays the State, that pays the
> school, that pays the contractor, that pays
> the paint . . . blah, blah, blah . . . that that
> is payment by the government of a false claim
> because the chemical manufacturer six or
> seven steps down the line commits fraud.[65]

Justice Scalia employs similar filler for a slightly different purpose, often stuttering slightly on the word "would" as he searches for words to criticize an advocate's position:

> JUSTICE SCALIA: I don't want to have to go
> before a jury as an employer all the time. I
> want—want a safe harbor. I don't even want to
> mess with people that might—that might be buy-
> ing a lawsuit, and you're telling me, well,
> you know, I can't help you. You have to go
> before a jury, say, if this person is close
> enough. Why can't we say members of family and
> fiancés? Would—would—would that be a nice
> rule?[66]

Just as interesting, though, are the question patterns that the Justices share. For example, without exception, all of the Justices frequently interrupt the advocates to introduce questions beginning with the word "well" when they have concerns about the advocate's argument. For Justice Alito it is often "well, let me ask"; for Justice Kagan, "well . . . I mean"; for Justice Sotomayor, "well, I think"; and so forth.[67] Indeed, the one gram "well" is one of the most predictive negative n-grams in every one of the Justice models. Advocates who have their arguments interrupted with a "well" question would do well to listen carefully and see if they can find a way to assuage the Justice's concerns.

C.  *Communication Between Justices*

Oral argument is the first opportunity the Justices have to get a sense of their colleagues' views and also their first chance to influence those views.⁶⁸ The highly predictive n-grams clearly show the important function oral argument serves as a platform for interjustice communication. At the center of that communication are two of the Court's most active participants in oral argument, Justices Breyer and Scalia. This is apparent, not primarily from their own predictive n-grams, but from those of their colleagues.

---

[65] Transcript of Oral Argument at 34, Allison Engine Co. v. United States *ex rel.* Sanders, 553 U.S. 662 (2008) (No. 07-214).

[66] Transcript of Oral Argument at 23, Thompson v. N. Am. Stainless, LP, 562 U.S. 170 (2011) (No. 09-291).

[67] *See, e.g.*, Transcript of Oral Argument at 6, United States Army Corps of Eng'rs v. Hawkes Co., 136 S. Ct. 1807 (2016) (No. 15-290) ("Well, let me just ask about how far removed it is.") (Alito, J.); Transcript of Oral Argument at 42, Baker Botts LLP. v. ASARCO LLC, 135 S. Ct. 2158 (2015) (No. 14-103) ("Well, it doesn't have to go be—I mean, I just don't understand what that means.") (Kagan, J.); Transcript of Oral Argument at 13, Chase Bank USA, N.A. v. McCoy, 562 U.S. 195 (2011) (No. 09-329) ("Well, I do think—I do think, counsel, that that major change doesn't have to be the way you describe it.") (Sotomayor, J.).

Among Justice Kennedy's most predictive negative n-grams, for example, is the two-gram "justic breyer," which he commonly uses to reference a past point raised by Justice Breyer as part of his own question. As just one example, in *United Haulers Association v. Oneida-Herkimer Solid Waste Management Authority*, Justice Kennedy framed his question to counsel for United Haulers as follows:

> JUSTICE KENNEDY: Well, I . . . is there a distinction between the question Justice Breyer put to you, the hypothetical of a municipal electricity company, and this case? In this case you have private haulers, you have private waste dumps at the end, you just have a public, a publicly owned and mandated processing center in the middle. It would be as if in the electric case you have private electric companies that generate the power, private electric companies that distribute the power, but they all have to go through a Government-owned transformer at the key. It seems that's the case you have here.⁶⁹

Across 691 cases in the study period, Justice Kennedy posed 171 such questions referencing Justice Breyer. The "justic breyer" n-gram's correlation with Justice Kennedy's vote for the party to whom it is asked suggests that Justice Kennedy finds Justice Breyer's questions both powerful and persuasive.

Justice Kennedy is no exception in this regard. References to other Justices' questions featured prominently in the models for all of the Justices, with every one of the Justices commonly referencing the questions of each of the other Justices. The table below shows the number of questions asked by each Justice referencing each other Justice during the study period. Because the composition of the court has changed over time and not all of the Justices have participated in an equal number of oral arguments together, the table depicts the number of questions asked per 100 cases that the Justices heard together.

---

⁶⁸ *See* Ryan C. Black, Timothy R. Johnson & Ryan J. Owens. *Chief Justice Burger and the Bench: How Physically Changing the Shape of the Court's Bench Reduced Interruptions during Oral Argument*, 43 J. of Sup. Ct. Hist. 83, 88 (2018); Ryan C. Black et al., *Emotions, Oral Arguments, and Supreme Court Decision Making*, 73 J. of Pol. 572, 574 (2011); Shapiro, *supra* note 57 at 530–31.

⁶⁹ Transcript of Oral Argument at 21, United Haulers Ass'n, Inc. v. Oneida-Herkimer Solid Waste Mgmt. Auth., 550 U.S. 330 (2007) (No. 05-1345).

| Referencing → <br> Speaking ↓ | Alito | Breyer | Ginsburg | Kagan | Kennedy | O'Connor | Rehnquist | Roberts | Scalia | Sotomayor | Souter | Stevens |
|---|---|---|---|---|---|---|---|---|---|---|---|---|
| Alito | 0 | 2 | 2 | 5 | 1 | 0 | 0 | 2 | 2 | 2 | 1 | 0 |
| Breyer | 8 | 0 | 7 | 7 | 7 | 6 | 3 | 7 | 15 | 6 | 6 | 5 |
| Ginsburg | 3 | 7 | 0 | 4 | 4 | 6 | 2 | 1 | 7 | 3 | 4 | 2 |
| Kagan | 12 | 12 | 10 | 0 | 7 | 0 | 0 | 10 | 15 | 6 | 0 | 0 |
| Kennedy | 13 | 11 | 9 | 14 | 0 | 9 | 6 | 10 | 13 | 2 | 7 | 6 |
| O'Connor | 0 | 1 | 1 | 0 | 1 | 0 | 0 | 0 | 1 | 0 | 0 | 0 |
| Rehnquist | 0 | 3 | 2 | 0 | 2 | 2 | 0 | 0 | 5 | 0 | 2 | 2 |
| Roberts | 9 | 11 | 5 | 6 | 8 | 0 | 0 | 0 | 6 | 6 | 6 | 11 |
| Scalia | 4 | 10 | 4 | 4 | 3 | 3 | 3 | 3 | 0 | 2 | 6 | 3 |
| Sotomayor | 12 | 24 | 9 | 11 | 7 | 0 | 0 | 5 | 16 | 0 | 0 | 4 |
| Souter | 6 | 18 | 11 | 0 | 7 | 5 | 4 | 4 | 17 | 0 | 0 | 9 |
| Stevens | 3 | 3 | 2 | 0 | 4 | 3 | 3 | 4 | 8 | 2 | 1 | 0 |

Justice Kennedy outpaces the other Justices in number of questions asked that reference another Justice. This fact may be attributable to Justice Kennedy's position as a swing vote in many cases, open to persuasion in either direction. As for the Justices who are referenced, Justices Scalia and Breyer are mentioned far more often in questions than are any of their fellow Justices. Indeed, other than Justice Alito, their questions are referenced more than twice as often as any other Justice. This is remarkable, but not surprising, and can be attributed at least in part to the greater number of questions asked by Justices Scalia and Breyer in comparison to their colleagues.

The data also show consistent across-the-aisle dialogue between the Court's liberals and conservatives. For example, Justice Scalia's questions are referenced more often in questions by Justices Souter, Sotomayor, and Kagan than questions by any of the Court's other conservatives. And Justice Kennedy references Justice Kagan in his questions more often than he references any of the Court's conservatives.

These across-the-aisle questions include attempts to persuade colleagues, recycled hypothetical questions, jokes, and subtle, or not-so-subtle, efforts to assist flailing advocates in answering tricky questions. The pattern in no way calls into question the long-standing division between the Court's liberals and conservatives. That divide is as strong as ever.[70] What the data do show, though, is encouraging—a consistent dialogue between the Court's liberal and conservative camps as the Justices engage all sides of the issues before them.

Conclusion

As the first study to leverage modern machine learning techniques to analyze Supreme Court oral argument dialogue, this Article breaks important new ground. Previous empirical studies of oral argument and judicial decision making have been limited to a small number of discrete, quantifiable attributes, many of them external to oral argument. Past work has examined, for instance, the strength of the connection between the Justices' votes and their ideological preferences; variation in the Justices' votes based on the legal subject matter at issue; and the Justices' tendency to ask more and longer questions to the parties they ultimately vote against.

This study breaks through past limitations, borrowing techniques from the fields of artificial intelligence and machine learning to analyze not only counts of questions and ideological leanings, but the actual oral argument dialogue itself—the content of the Justices' questions. In so doing, the Article offers an important new window into aspects of oral argument that have long resisted empirical study, including the Justices' individual questioning styles, how each expresses skepticism, what inter-Justice dialogue looks like for each Justice, and which of the Justices' questions most drive oral argument.

Still, although this Article offers the most in-depth empirical analysis of Supreme Court oral argument dialogue to date, it is only an initial foray. It necessarily leaves many avenues unexplored and ripe for future analysis. What sorts of considerations are most important to the Justices? How do those considerations vary across case types? Whose questions most drive the final decision? The Article also leaves for the future ancillary questions such as how oral argument has evolved over time, whether and how the Justices' oral argument styles change in publicly salient cases, and what role oral argument plays in the formation of voting coalitions among the Justices. In short, this Article demonstrates a powerful set of new techniques for exploring oral argument, but much work still remains to be done. The author hopes that, beyond the important new insights it offers into oral argument, this Article will also spur increased interest in the application of machine learning as a tool for legal scholarship more generally, particularly for those questions ill-suited to more traditional methodologies.

---

[70] *See How America's Supreme Court Became so Politicised*, The Economist (Sept. 15, 2018), https://www.economist.com/briefing/2018/09/15/how-americas-supreme-court-became-so-politicised; Neal Devins & Lawrence Baum, The Company They Keep: How Partisan Divisions Came to the Supreme Court (forthcoming 2019) (manuscript at 239).